\documentstyle[12pt]{article}
\begin{document}
\input epsf
\begin{titlepage}
\begin{center}
\today     \hfill    WM-00-114 

\vskip .5in

{\large \bf  Orthogonal U(1)'s, Proton Stability and Extra Dimensions}

\vskip 0.5in

Alfredo Aranda\footnote{fefo@physics.wm.edu} and 
Christopher D. Carone\footnote{carone@physics.wm.edu}

{\em Nuclear and Particle Theory Group\\
Department of Physics, College of William and Mary \\
Williamsburg, VA 23187}

\vskip 0.5in

\end{center}

\vskip .1in

\begin{abstract}
In models with a low quantum gravity scale, one might expect that all 
operators consistent with gauge symmetries are present in the low-energy 
effective theory.  If this is the case, some mechanism must be present 
to adequately suppress operators that violate baryon number. Here we 
explore the possibility that the desired suppression is a consequence 
of an additional, spontaneously-broken, non-anomalous U(1) symmetry 
that is orthogonal to hypercharge.  We show that successful models can be 
constructed in which the additional particle content necessary to cancel 
anomalies is minimal, and compatible with the constraints from precision 
electroweak measurements and gauge unification.  If unification is
sacrificed, and only the new U(1) and its associated Higgs fields live in 
the bulk, it is possible that the gauge field zero mode and first few 
Kaluza-Klein excitations lie within the kinematic reach of the Tevatron.  
For gauge couplings not much smaller than that of hypercharge, we show that 
these highly leptophobic states could  evade detection at Run I, but be 
discovered at Run II.  Our scenario presents an alternative to the 
`cartographic' solution to baryon number violation in which leptons and 
quarks are separated in an extra dimension.
\end{abstract}

\end{titlepage}

\newpage
\renewcommand{\thepage}{\arabic{page}}
\setcounter{page}{1}
\section{Introduction} \label{sec:intro} \setcounter{equation}{0}

It is a general principle of effective field theory that one
should include all operators consistent with symmetry constraints 
when constructing a low-energy effective Lagrangian~\cite{georgi}.  
Such operators are suppressed by powers of the ultraviolet cutoff, so 
that each has the appropriate mass dimension, and multiplied by 
coefficients that parameterize the unknown physics relevant at higher 
energy scales.  When this approach is applied to models with a low 
quantum gravity scale~\cite{add}, one obtains a multitude of 
phenomenological disasters, unless specific mechanisms are invoked to 
suppress contributions to processes that are suppressed or absent in the 
standard model~\cite{fv}.  In this paper, we consider the possibility 
that baryon-number-violating operators are present generically in 
such theories~\cite{kane}, but are suppressed by an additional, non-anomalous, 
spontaneously-broken U(1) gauge symmetry that is orthogonal to 
hypercharge~\cite{carmur2}.  We will argue that the natural scale for 
the breaking of this symmetry is ${\cal O}(1)$~TeV, so that our scenario 
may have testable consequences at the Fermilab Tevatron, or at the next 
generation of collider experiments.

We focus on baryon number violation since it is by far the most
dangerous of nonstandard model processes.  Even if the Planck
scale has its conventional value $M_{Pl}\approx 10^{19}$~GeV,
the most general set of Planck-suppressed, baryon-number-violating 
operators lead to proton decay at a rate that is much too fast,
unless there is some additional parametric suppression.  For example,
the superpotential operator $(Q_1 Q_{1,2})Q_2 L_i/M_{Pl}$ must be
suppressed by an additional factor of ${\cal O}( 10^{-6})$ to avoid
conflict with the proton lifetime bounds from SuperKamiokande~\cite{sk}.
For a high Planck scale, this additional suppression factor can
originate from the same sequential breaking of flavor symmetries that 
may account for the smallness of the Yukawa couplings of the first two 
generations~\cite{s33}.  However, if $M_{Pl}$ is in the $1-100$~TeV range, 
which can be the case in models with extra spacetime dimensions compactified
at the TeV-scale, then a much higher degree of suppression is required.
We will show that a flavor-universal U(1) gauge symmetry, isomorphic to 
baryon number on the standard model particle content and spontaneously 
broken only slightly above the weak scale, is sufficient to avoid any
phenomenological problems stemming from baryon-number-violating 
operators.

It is worth stressing that there are probably many possible ways of
suppressing or eliminating proton decay in 
theories with a low Planck scale.  One elegant suggestion made
by Arkani-Hamed and Schmaltz is that quarks and leptons may be
localized at different points in an extra dimension, so that proton
decay operators are suppressed by the tiny overlap of the quark and 
lepton wave functions~\cite{ahs}.  The approach that we consider here is
complementary in that it applies also to the case when quarks and
leptons are fixed to a single brane, with no separation.  No doubt, 
this possibility has met considerable interest in the recent
literature~\cite{br}.

There is some relationship between the present work and earlier 
papers on the possibility of gauged baryon number, in which the 
scale of spontaneous symmetry breaking was taken {\em below} 
$M_Z$~\cite{carmur2,carmur1,dav,ac}.  While the proton decay issue was 
discussed in Ref.~\cite{carmur2}, the model used as a basis for the 
argument is now excluded at above the $95\%$ confidence level from bounds 
on the electroweak $S$ parameter -- the model required a fourth chiral 
generation to cancel gauge anomalies.  Other possibilities for anomaly 
cancellation discussed in the first version of Ref.~\cite{carmur1} are excluded
by $S$, and are also inconsistent with gauge coupling unification.  
Here we will present a supersymmetric model that is consistent with 
unification (in the case where all gauge and Higgs fields live in the 
bulk~\cite{aa,ddg}) as well as the  anomaly-cancellation constraints.   The 
required extra matter is chiral under the full gauge group, but vector-like 
under the standard model gauge factors, so that the  $S$ parameter bound 
may be avoided.  The extra matter fields get masses of order the U(1) breaking 
scale $\Lambda_B$, which in principle could be decoupled from the weak 
scale.  We suggest, however, that a natural possibility for generating 
$\Lambda_B$ is a radiative breaking scenario that relates this scale to 
the scale of supersymmetry breaking.  In this case, the new physics 
we introduce becomes relevant for TeV-scale collider experiments.

One of the distinctive features of the $Z'$ boson in the class of 
models we consider is its natural leptophobia.  While it may be tempting 
to think that a model with gauged baryon number is leptophobic by design, 
it is not hard to see that this statement is patently false.  Generically, 
any additional U(1) symmetry will mix with hypercharge via
the kinetic interaction
\begin{equation}
{\cal L} =- \frac{1}{2} c_B F^{\mu\nu}_Y F_{\mu\nu}^{new} \,\,\, ,
\label{eq:kinmix}
\end{equation}
which is not forbidden by any symmetry of the low-energy theory.  Even 
if $c_B$ is identically zero at the ultraviolet cutoff of the
theory $M_{Pl}$, it will be renormalized at one loop by all particles 
that carry both hypercharge and the additional U(1) charge,
so that $c_B(\mu)\neq 0$ for $\mu < M_{Pl}$.  The class of models that 
we consider here have the property that $c_B(M_{Pl})=0$, and in addition
\begin{equation}
Tr(BY)=0 \,\,\, ,
\end{equation}
where $B$ and $Y$ are the baryon number and hypercharge matrices, 
and the trace sums over all fields in the theory.  It is in this
sense that we say the additional U(1) is orthogonal to hypercharge.
Such orthogonal U(1)'s are known to arise in string theory~\cite{dkm}, 
though we will not commit ourselves to any specific string-theoretic embedding.
The constraint $Tr(BY)=0$ assures that the mixing parameter $c_B(\mu)$ 
remains zero until the heaviest particle threshold is crossed.  In our 
models, the heaviest particle threshold includes all the nonstandard 
particles introduced to cancel anomalies; thus the running of $c_B(\mu)$ 
begins after the exotic  states are integrated out, and hence is controlled 
{\em solely} by the standard model particle content.  This gives our 
phenomenological analysis a high degree of model independence:  a similar
model with different nonstandard matter content would have identical $Z'$ 
phenomenology.\footnote{For $Z'$ models that suppress proton decay
and have a different phenomenology, see Ref.~\cite{erler}.}    

It is worth stressing that the leptophobia of the $Z'$ in this
model (as well as the leptophobia of its Kaluza-Klein excitations)
is quite robust.  For example, one might think that the $Z'$ could
be made less leptophobic by taking the scale $\Lambda_B$ to be
high (so that $c_B(\mu)$ would have a greater distance to run).
However, this possibility is inconsistent with the assumption
that (a) the $Z'$ zero mode is phenomenologically relevant and (b) the model
is consistent with unification.  Since we don't know the string 
normalization of the new U(1) gauge coupling, we only require that it 
not differ wildly in strength from hypercharge at low energies.  For 
a $Z'$ with mass $M_B<1$~TeV, and coupling $g_B \mbox{\raisebox{-1.0ex} 
{$\stackrel{\textstyle ~<~} {\textstyle \sim}$}} g_Y$, the 
associated symmetry breaking scale $M_B/g_B$ cannot be arbitrarily 
high.  Since this is also the scale of the exotic matter content, 
$c_B(\mu)$ cannot run over very large intervals.  If one takes
$g_B$ to be smaller, the scale at which running begins is pushed
up, but $c_B(\mu)$ runs more slowly due to the reduced coupling.
We study this effect quantitatively in Section~3.

Finally, if one is willing to sacrifice simple power-law unification,
as in the original scenario of Arkani-Hamed, Dimopoulos and 
Dvali~\cite{add}, then it is possible to consider a scenario where only
gravity and the additional U(1) may propagate into the extra
dimensional bulk space.  What is interesting about this
possibility is that strongest bounds on the compactification
scale come solely from the effects of the new U(1).  As a consequence,
the $Z'$ and its Kaluza-Klein (KK) excitations may be brought within the 
kinematic reach of the Tevatron.  We show that for gauge couplings not much 
smaller than that of hypercharge, the $Z'$ and its first few KK modes could 
remain invisible at Run I of the Tevatron, but be discerned easily at 
Run II.  For this model, the ability of a collider experiment to
probe weak couplings is as important as mass reach; we show that the 
enhanced luminosity of Run II could allow the Tevatron to probe a 
significant region of the model's parameter space.

In the next section, we highlight the points discussed above by
presenting a concrete example.  We do not view this model as
unique, but rather as a representative example of a class
of orthogonal U(1) models that have similar low-energy physics.  
In Section~3 we discuss the low-energy phenomenology of our scenario, and 
in the final section present our conclusions.

\section{A Model} \setcounter{equation}{0}

The gauge group is that of the standard model with an additional 
U(1) factor:
\begin{equation}
G = SU(3) \times SU(2) \times U(1)_Y \times U(1)_B  \,\,\, .
\end{equation}
We normalize the gauge coupling $g_B$ such that all standard model 
quarks have charge $1/3$, while all leptons and standard model Higgs 
fields have charge $0$; these are the conventional charge assignments 
for baryon number in the standard model.  Gauging this symmetry requires 
the introduction of exotic matter to cancel chiral gauge anomalies, as 
well as additional Higgs fields to spontaneously break the symmetry
and avoid long-range forces.  The aim of this section is to show that 
this can be done in a relatively simple way, consistent with a number 
of important phenomenological constraints.  In particular, we show that 
exotic matter can be chosen such that the model (1) is consistent with 
gauge unification, (2) is anomaly free,  (3) suppresses proton decay 
sufficiently, (4) has no unwanted stable colored or charged states,
and (5) has a mechanism for giving the exotic matter mass.  We present
the model by considering these issues systematically:

{\em Gauge Unification}.  We would like our model to be consistent 
with power-law unification~\cite{ddg}, at least in the case where all 
the gauge and Higgs fields are allowed to propagate into the extra-dimensional 
space.  Since the string normalization of the additional U(1) is
uncertain~\cite{dfm}, we seek to preserve unification of the ordinary
standard model gauge factors while allowing $g_B$ to assume values 
at low energies that do not differ wildly from that of hypercharge. We 
therefore require that the exotic matter fields fall in complete SU(5) 
representations.  While there are of course other possibilities~\cite{gu}, 
this is the simplest.  We introduce an extra generation that is vector-like 
under the standard model gauge factors but chiral under U(1)$_B$:
\begin{equation}
\begin{array}{cc}
\left. \begin{array}{c}
Q_L \\ 
U_R \\ 
D_R \\
\end{array} \right\} b_Q
&
\left. \begin{array}{c}
\overline{Q}_R \\ 
\overline{U}_L \\ 
\overline{D}_L \\
\end{array} \right\} b_{\bar{Q}}
\\ \\
\left. \begin{array}{c} 
L_L \\ 
E_R \\ 
N_R \\
\end{array} \right\} b_L
&
\left. \begin{array}{c} 
\overline{L}_R \\ 
\overline{E}_L \\ 
\overline{N}_L \\
\end{array} \right\} b_{\bar{L}}
\end{array} \,\,\, .
\label{eq:fields}
\end{equation} 
Although we assume supersymmetry, we show only the fermionic
components above. The overlines indicate Dirac adjoints, and
the $b$'s represent the U(1)$_B$ charges, yet to be specified.
(Four distinct U(1)$_B$ charges is the smallest number we found that 
could produce a viable model.)  The charges under the standard model 
gauge factors for fields in the first column are precisely the same 
as those of fields in an ordinary standard model generation; the only 
exception is $N_R$ which is a standard model singlet.  The fields in the 
second column have conjugate standard model charges so that, for example,
$\overline{Q}_R Q_L$ would be invariant if $b_Q+b_{\bar{Q}}=0$.
As we will see below, our choices for the $b_i$ are such that
all the fields in Eq.~(\ref{eq:fields}) obtain masses of order
the U(1)$_B$ breaking scale.

{\em Anomaly Cancellation} We now aim to restrict the $b_i$
so that the model is free of gauge anomalies.  We first note that
triangle diagrams involving only standard model gauge factors
remain vanishing since the additional matter is introduced in
complete generations.  We therefore must consider 
anomalies of the form U(1)$^3_B$, $G_{SM}$U(1)$^2_B$ and $G_{SM}^2$U(1)$_B$,
where $G_{SM}$ represents any of the standard model group factors.
Given the tracelessness of the non-Abelian generators, this
reduces the relevant anomalies to the set:  U(1)$_Y$U(1)$_B^2$, 
SU(3)$^2$U(1)$_B$, SU(2)$^2$U(1)$_B$, U(1)$_Y^2$U(1)$_B$, and U(1)$_B^3$. 
It is easy to see that the SU(3)$^2$U(1)$_B$ anomaly vanishes
since all colored matter with the same U(1)$_B$ charge comes in
groups with equal numbers of left- and right-handed fields.
The same can be said of the U(1)$_B^3$ anomaly, since the 
additional $N_{L,R}$ states assure that the exotic `leptons' with the
same U(1)$_B$ charge again come in equal numbers of left- and
right-handed fields.  Finally, we can dispense with the 
U(1)$_Y$U(1)$_B^2$ anomaly by noting that every group of particles
with the same U(1)$_B$ charge separately satisfies $Tr(Y)=0$.
The remaining two anomaly cancellation conditions,
SU(2)$^2$U(1)$_B$ and U(1)$_Y^2$U(1)$_B$, give exactly 
the same constraint
\begin{equation}
3 \Delta_Q + \Delta_L = -3 \,\,\, ,
\label{eq:anomcon}
\end{equation}
where we have defined
\begin{equation}
\Delta_Q = b_Q + b_{\bar{Q}} \,\,\,\,\,\mbox{ and } \,\,\,\,\,
\Delta_L = b_L + b_{\bar{L}} \,\,\, .
\end{equation}
Given the charges defined in Eq.~(\ref{eq:fields}), we impose 
Eq.~(\ref{eq:anomcon}) to render our theory free of anomalies.

Notice that $-\Delta_Q$ and $-\Delta_L$ also represent the charges of 
Higgs fields that we require to give the exotic matter fields masses
when U(1)$_B$ is spontaneously broken.  The most economical exotic
Higgs sector is obtained by setting
\begin{equation}
\Delta_Q  = \pm\Delta_L  \,\,\, .
\label{eq:smallhiggs}
\end{equation}
Then all the desired mass terms may be formed by introducing
a single pair of Higgs fields
\begin{equation}
S_B  \,\,\,\,\, \mbox{ and } \,\,\,\,\, S_{\bar{B}} \,\,\, ,
\end{equation}
with charges $+\Delta_Q$ and $-\Delta_Q$, respectively. This is
the minimal possibility, since, as in the minimal supersymmetric standard
model (MSSM), a vector-like pair of Higgs superfields is required to 
avoid additional anomalies.  The choice of Eq.~(\ref{eq:smallhiggs}) together
with the constraint Eq.~(\ref{eq:anomcon}) implies that 
either
\begin{equation}
\Delta_Q = \Delta_L= -3/4 \,\,\,\,\, \mbox{ or } 
\,\,\,\,\, \Delta_Q= -\Delta_L= -3/2
\,\,\, .
\end{equation}
The remaining freedom to choose exotic U(1)$_B$ charges will be important 
in satisfying the other phenomenological constraints below. 

{\em Proton decay}.  If our additional U(1) symmetry were unbroken, then it 
would be clear that all operators contributing to proton decay would 
be exactly forbidden.  When the symmetry is spontaneously broken, the form 
of baryon-number-violating operators in the low-energy effective theory
depends on the charge assignment of the Higgs fields which break U(1)$_B$, 
as well as on the size of their vacuum expectation values (vevs).   Let us 
work in the very low-energy limit, below the scales of extra dimensions, exotic
matter, and supersymmetry breaking, which we will take to be $\sim 1$~TeV
universally for the purposes of the present argument.  In this effective
nonsupersymmetric theory, operators that could contribute to proton decay 
have the form~\cite{carmur2}
\begin{equation}
{\cal O} = q^k \ell^m \chi^n  \,\,\, ,
\label{eq:bfirst}
\end{equation}
where $q$ and $\ell$ represent generic quark and lepton fields, respectively, 
and $\chi$ represents the vev of either $S_B$ or $S_{\bar{B}}$.  Here we
have suppressed both the Dirac structure of the operator and the standard
model gauge indices for convenience.  First, we  note that since the 
lepton electric charge is integral, $k$ must be a multiple of $3$, 
{\em i.e.} $k=3 p$.  It follows that the baryon number of $q^k\equiv q^{3p}$ 
is $p$, which is an integer.  On the other hand, this
must be compensated by the baryon number of $\chi$, which is either $\pm 3/2$ 
or $\pm 3/4$, given the charges of the $S_B$ fields already discussed.  
Thus we conclude that the operators represented by Eq.~(\ref{eq:bfirst}) 
must be of the form
\begin{equation}
(q^9 \chi^2)^r \ell^m   \,\,\,\,\, \mbox{ or } \,\,\,\,\,
(q^9 \chi^4)^r \ell^m  \,\,\, ,
\label{eq:bigops}
\end{equation}
where $r$ and $m$ are integers.  The point is simple: the fact that the 
possible symmetry breaking `spurions' have fractional U(1)$_B$ charges forces 
the baryon-number-violating operators to contribute to no less than 
$\Delta B = 3$ transitions.  This renders our model safe from proton decay 
as well as $N$-$\overline{N}$ oscillations.  The operators in
Eq.~(\ref{eq:bigops}) are suppressed by high powers of mass scales that are
either $1$~TeV or $M_{Pl}$, and thus are unlikely to have any observable 
effects on stable matter at low energies.

{\em Avoiding Stable Charged Exotic Matter.}  We will now further 
restrict our charge assignments $b_i$ to assure that we have no stable heavy 
states that are charged under any of the standard model gauge factors.
This allows us to evade bounds on stable charged matter from searches
for anomalously heavy isotopes in sea water~\cite{rpp00}. In 
both the exotic lepton and quark sectors separately, it is always possible to 
choose Yukawa couplings such that one exotic state is lightest, and ordinary 
weak decays to this state are kinematically allowed.  For example, the exotic 
lepton superpotential couplings (in terms of left-handed chiral
superfields)
\begin{equation}
W \supset L\bar{L}S_{\bar{B}} + (E\bar{E}+N\overline{N})S_B 
+ (LE+\bar{L}\bar{N})H_D + (\bar{L}\bar{E}+LN)H_U
\end{equation}
lead to mass terms of the form
\begin{equation}
(\begin{array}{cc} \overline{e}_H & E \end{array})
\left(\begin{array}{cc} M_1 & m_2 \\ m_1 & M_2 \end{array}\right)
\left(\begin{array}{c} e_H \\ \overline{E}\end{array}\right) +
(\begin{array}{cc} \overline{\nu}_H & N \end{array})
\left(\begin{array}{cc} M_1 & m_4 \\ m_3 & M_3 \end{array}\right)
\left(\begin{array}{c} \nu_H \\ \overline{N}\end{array}\right) \,\,\, ,
\end{equation}
where the $M_i$ are masses of order the U(1)$_B$ breaking scale,
while $m_i$ are of order the weak scale.   Here we have
written the component superfields in the doublets $L$ ($\overline{L}$)
as $\nu_H$ ($\overline{\nu}_H$) and  $e_H$ ($\overline{e}_H$).
Clearly one has the freedom to arrange for the lightest 
exotic lepton state to be neutral.  For example, for
the specific choice $M_1=M_2=M_3$, $m_1=m_2$ and $m_3=m_4$, the lightest 
charged state has mass $M-m_1$ while the lightest neutral state $M-m_3$; we
therefore could take $m_1<m_3$. In the exotic quark sector, the
lightest state is charged and colored, so some additional mechanism
must be provided to assure it decays to ordinary particles.  Since 
we are working in the context of models in which the Planck scale is low,
we can make the lightest exotic quark unstable by considering possible
higher-dimension operators, allowed by the symmetries of the theory
and suppressed by the cutoff.   As there is some freedom in how we may
accomplish this, let us restrict our subsequent discussion to a 
specific example.    Let us choose the charge assignment in 
which $\Delta_Q=-3/2$.  The choice $b_Q=-2/3$ and $b_{\bar{Q}}=-5/6$
is consistent with this condition, and also allows the 
superpotential operator
\begin{equation}
\frac{1}{M_{Pl}} q\, q\, Q\, \ell \,\,\, ,
\label{eq:hdimop}
\end{equation}
where lower-case superfields are those of the standard model.  This operator
allows for three-body decays for the lightest exotic quark field  (for example,
to a normal lepton and two squarks).  Even if the superpartners are
heavy, so that this decay is not kinematically allowed, one can obtain
a four-fermion operator by ``dressing'' Eq.~(\ref{eq:hdimop}) with
a gaugino exchange.  In this case, the decay proceeds to two quarks
and a lepton, with a width of order
\begin{equation}
\Gamma \sim \frac{1}{64 \pi^3}\left(\frac{1}{16\pi^2}\right)^2
\left(\frac{M_Q}{M_{Pl}}\right)^2 M_Q  \,\,\, .
\end{equation}
The first factor is from three-body phase space, the second
from the fact that the amplitude occurs at one-loop, and the rest
follows from dimensional analysis.  The lightest exotic quark decays
well before nucleosynthesis providing that $M_{Pl}<10^{13}$~GeV; this
is not a problem in our scenario.  Note that the
charge assignments $b_Q=-2/3$ and $b_{\bar{Q}}=-5/6$ assure that
potentially dangerous mass mixing terms like $q\overline{Q}$,
and $Q H_D d$ have U(1)$_B$ charges of $-1/2$ and $-1$, respectively.  
Since this is not an integral multiple of $3/2$ (the magnitude
of the exotic Higgs' U(1)$_B$ charges) such operators are forbidden
by the gauge symmetry.  We will adopt the present choice of $b_Q$ and
$b_{\bar{Q}}$ for the subsequent discussion.  However, the reader
should keep in mind that other possible assignments may 
render the exotic matter unstable, given the presence of 
higher-dimension operators at the relatively low cutoff of the 
theory.

{\em Orthogonality}.  The only charges we have not yet fixed are
$b_L$ and $b_{\bar{L}}$, which have been constrained such that
$b_L + b_{\bar{L}}= 3/2$.   Since we wish to restrict our discussion
to models that satisfy $Tr(BY)=0$, we fix our remaining degree of
freedom by imposing this constraint.  It is straightforward
to check that $Tr(BY)= 9\cdot\frac{1}{3}\cdot(2\cdot \frac{1}{6}
+\frac{2}{3}-\frac{1}{3})=2$ for the ordinary matter, where the 
overall factor of $9$ is the multiplicity due to color and number of
generations.  For the exotic matter, the quark fields
contribute $Tr(BY)=3\cdot(b_Q-b_{\bar{Q}})\cdot(2\cdot \frac{1}{6}
+\frac{2}{3}-\frac{1}{3}) = 1/3$ given our previous choice of 
$b_Q=-2/3$ and $b_{\bar{Q}}=-5/6$.  We now choose
$b_L=4/3$ and $b_{\overline{L}}=1/6$.  The exotic lepton
contribution is then $Tr(BY)=(b_L-b_{\overline{L}})(2\cdot[-\frac{1}{2}]
-1)=-7/3$.  Hence, the orthogonality of U(1)$_B$ and hypercharge
is maintained.  Notice that our choice for $b_L$ and $b_{\bar{L}}$
is such that no dangerous mass mixing terms between exotic
and standard model leptons are generated after U(1)$_B$ is
spontaneously broken.  Now that all our charges have been
fixed, we summarize them here for convenience:
\begin{equation}
\begin{array}{ll}
b_Q=-2/3  & b_{\bar{Q}}=-5/6  \\
b_L= \,\,\,\,\, 4/3   & b_{\overline{L}}= \,\,\,\,\, 1/6
\end{array} \,\,\, .
\end{equation}

{\em Symmetry Breaking}.  It is customary in model building to 
avoid discussing the origin of symmetry breaking scales, given the 
model-dependence that this issue often entails.  Here we only aim 
to emphasize that the scale of U(1)$_B$ breaking may be tied quite 
naturally to the scale of supersymmetry breaking.  This point is 
worth mentioning given that we have constructed our model specifically 
to allow for the decoupling of the nonstandard sector, to avoid bounds 
from precision electroweak measurements. One way in which the supersymmetry 
breaking and U(1)$_B$ scale may be related is if the potential for the 
nonstandard Higgs fields $S_B$ and $S_{\bar{B}}$ develops its vacuum 
expectation value as a consequence of a soft scalar squared mass running
negative, the analog of the radiative breaking scenario in
the MSSM.  This scenario can be implemented in the present context since 
the exotic Higgs fields couple to a sector of new matter fields with 
large Yukawa couplings.  The exotic Higgs fields have the superpotential 
coupling
\begin{equation}
W=\mu_s S_B S_{\bar{B}},
\end{equation}
the analog of the $\mu$ term in the MSSM.  Introducing
soft supersymmetry breaking masses, and $D$-terms, the scalar
potential for the exotic Higgs fields is given by
\begin{eqnarray}
V &=& \frac{1}{2} (\mu^2_s + m_B^2) (s_B^2 +p_B^2)
+ \frac{1}{2} (\mu^2_s + m_{\bar{B}}^2) (s_{\bar{B}}^2 +p_{\bar{B}}^2)
\nonumber \\
&+&\mu_s B_s (s_B s_{\bar{B}}-p_B p_{\bar{B}})
+\frac{9}{32} g_B^2 (s_B^2 +p_B^2-s_{\bar{B}}^2-p_{\bar{B}}^2)^2  \,\,\, ,
\end{eqnarray}
where $s_{B,\bar{B}}$ and $p_{B,\bar{B}}$ represent the scalar and 
pseudoscalar components of each of the fields, and $m_B$, $m_{\bar{B}}$, and
$B_s$ are soft, supersymmetry-breaking masses.   It is
straightforward to show that this potential has stable (local)
minima in which one scalar squared mass is negative and both 
$S_B$ and $S_{\bar{B}}$ acquire vacuum expectation values.
For example, for the parameter choice $g_B=0.3$, $\mu_s =1$~TeV
$B_s=-1$~TeV, $m_B^2=-1.48$~TeV$^2$, and $m_{\bar{B}}^2=2.81$~TeV$^2$,
we find the vevs 
\[
\langle s_B \rangle = 3\mbox{ TeV}
\,\,\,\,\,\,
\langle s_{\bar{B}}\rangle =1\mbox{ TeV},
\]
the scalar squared masses
\[
0.99\mbox{ TeV}^2 \,\,\,\,\,\, 4.37\mbox{ TeV}^2 \,\, ,
\]
and the pseudoscalar squared mass
\[
3.33\mbox{ TeV}^2 \,\,\, . 
\]
These are acceptable values. Another possible
form for the potential is that of the next-to-minimal
supersymmetric standard model, in which both the
ordinary $\mu$ parameter and the parameter $\mu_s$
could have a common origin, the vev of a singlet
field.   We will not study the issue of possible potentials 
any further here, though such an investigation
would be required if experimental evidence for the
model became available.

\section{Phenomenology} \label{sec:bounds} \setcounter{equation}{0}

In this section, we explore the $Z'$ phenomenology of our model.
We will assume for simplicity that the scale of exotic matter,
$\Lambda_B$, and of superpartner masses is $1$~TeV.  The compactification
scale, which we call $\Lambda$ below, is a free parameter.  In the
case where all non-chiral matter (i.e. the Higgs and gauge fields)
are allowed to propagate in the bulk, we require $\Lambda$ to be
greater than a few TeV, to satisfy the constraints from precision
electroweak measurements~\cite{pem}.  In this case, the phenomenology
that we study is that of the new zero mode gauge field.  However, we will 
also consider the (non-unifiable) possibility that only U(1)$_B$ lives 
in the bulk, in which case the bounds on $\Lambda$ 
are substantially weakened.  For this choice, the $Z'$ zero mode and
first few KK excitations become relevant at planned collider
experiments, and will be the focus of our discussion.  For concreteness,
we perform our numerical analysis in the case of one extra 
dimension.\footnote{Of course, gravity also lives in the bulk.
Our model does not preclude the possibility that gravity
propagates in a larger number of dimensions than U(1)$_B$.}
For more than one extra dimension, the sums involving the KK modes
are divergent and must be regulated by some additional,
string-theoretic mechanism.  We restrict ourselves to one extra
dimension to avoid this model-dependent issue; however, the reader
should keep in mind that our bounds on the U(1)$_B$ KK modes may
be overestimates if there is a mechanism, {\em e.g.} brane
recoil effects~\cite{bre}, that suppresses the KK couplings.

One of the interesting properties of this class of models, regardless
of which case we consider, is the strong leptophobia of the $Z^{\prime}$
and its KK excitations.  Given our assumption of a vanishing kinetic 
mixing parameter, $c_B$, at the string scale, $c_B$ remains 
vanishing down to the scale of exotic matter, since $Tr(BY)=0$.
At lower scales, the exotic states are integrated out of the theory,
and the orthogonality constraint is no longer satisfied.  With
our choice of energy scales, $c_B$ remains small down to the
$Z'$ mass, so we may treat Eq.~(\ref{eq:kinmix}) as a perturbative 
interaction.  Thus, the Feynman rule for the $Z'$-hypercharge vertex is 
given by
\begin{equation} \label{feynmanrule}
-i \, c_B \, \left( p^2g^{\mu \nu} - p^{\mu}p^{\nu} \right) \, .
\end{equation}
Since we assume that the scale of superpartner masses is the same as 
the scale of exotic matter, we evaluate the non-supersymmetric running 
of $c_B$; at one-loop we obtain the renormalization group equation (RGE)
\begin{equation} \label{rge}
\mu \frac{\partial}{\partial \mu} c_B = 
-\frac{1}{3 \pi} \sqrt{\alpha_Y \alpha_B} \left[\frac{5}{6} N_u -
\frac{1}{6}N_d \right] \,\,\, ,
\end{equation}
where $N_u$ and $N_d$ are the number of standard model up-type and 
down-type quarks propagating in the loop.  This RGE is solved subject to 
the boundary condition $c_B(\Lambda_B)=0$, for the reasons described above.
Notice that the running of $c_B$ is controlled entirely by the standard model
particle content, since these are the only fields relevant below
the scale $\Lambda_B$.  Thus, our analysis is independent of the
specific exotic sector introduced to cancel anomalies.

We may now consider the phenomenology of the model by determining
bounds in the $M_B$-$\alpha_B$ plane.  We will assume  $M_B > m_{top}$ 
(which was not studied in Refs.~\cite{carmur2,carmur1}) and first 
consider the case in which all non-chiral superfields live in
the bulk.  For most of the mass range of
interest, the $Z'$ will be sufficiently heavier than the $Z$ so
that the most stringent bounds are obtained from direct collider
searches.  We consider the limits on  $Z'$'s decaying
to dijets and dileptons at the Fermilab Tevatron Collider:

\begin{figure}  
\centerline{\epsfysize=2.75 in \epsfbox{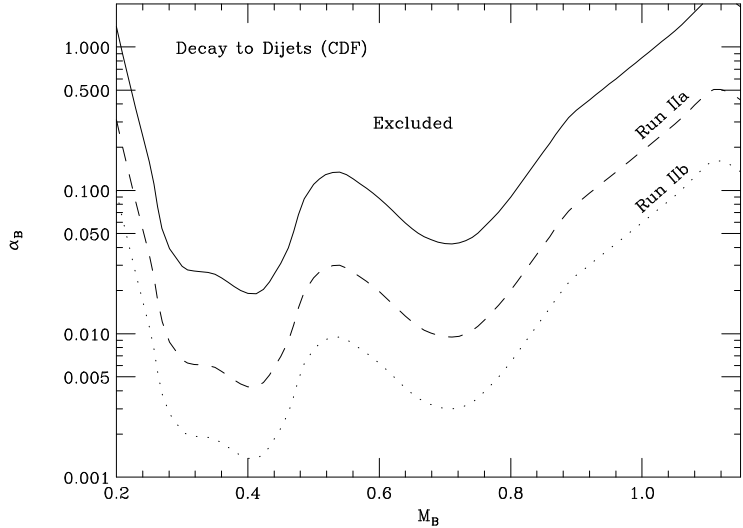}}
\caption{Bound on $\alpha_B$ from the cross section times
branching fraction to dijets. The solid line corresponds
to the bound obtained from Run I with a Luminosity of 
106 pb$^{-1}$. The dashed line corresponds to a luminosity
of 2 fb$^{-1}$ for Run IIa and the dotted line to a
luminosity of 20 fb$^{-1}$ for Run IIb.}
\label{figure1}
\end{figure}

{\em Decays to Dijets}.  The CDF Collaboration has placed
bounds on narrow resonances decaying to dijets in $p\bar{p}$ collisions
at $\sqrt{s}=1.8$~TeV~\cite{abe1}. They present the $95\%$ C.L. 
upper limits on cross section times branching ratio as a function of 
the $Z'$ mass in the range $0.2-1.15$~TeV.  Since the kinetic
mixing effects are small (as we will see below), the branching
fraction to dijets in our model is nearly 100\%; thus we compare
the CDF bounds to the $Z'$ production cross section in our model, which 
we estimate using the narrow width approximation:
\begin{equation} \label{eq:dijet}
\sigma(p\bar{p} \rightarrow Z^{\prime} \rightarrow dijets) = 
\frac{4 \pi^2}{9} \frac{\alpha_B}{s} \int dy \sum_{i,j}
f^{p}_i(y,\sqrt{s},M_B) f^{\bar{p}}_j(y,\sqrt{s},M_B)
\, .
\end{equation}
Here $y$ is the rapidity, $\sqrt{s}$ is the center of mass energy, and
$f^p$ ($f^{\bar{p}}$) represents the appropriate parton distribution functions
for $p\bar{p}$ collisions.  Using the CTEQ 4M structure functions~\cite{pdf} 
at $\sqrt{s} = 1.8$~TeV for our numerical analysis, we obtain a bound 
on $\alpha_B(M_B)$ as a function of $M_B$, shown in Fig.~\ref{figure1}.
The solid line corresponds to the Run I luminosity (${\cal L}$) of 
$\sim$0.1 fb$^{-1}$ and is the strongest bound on the model.  We also
estimate the ability of the Tevatron to probe additional parameter
space at Run~II.  Note that the shape of the excluded region
in Fig.~\ref{figure1} depends on a detailed analysis of both
statistical and experimental systematic uncertainties; the latter are 
difficult to extrapolate with precision to Run~II.  Therefore, we rely
instead on the observation that statistical and systematic 
uncertainties generally both scale as $\sqrt{{ \cal L}}$ ({\em i.e.} the
systematic uncertainties can be reduced by higher statistics). Thus,
we make a simple extrapolation, scaling the bound from Run I down
by $\sqrt{{\cal L}_I}/\sqrt{{ \cal L}_{II}}$ using the expected 
luminosities at Run IIa and Run IIb, 2 and 20 fb$^{-1}$ respectively; 
this yields the two other curves shown in Fig.~\ref{figure1}. 
We see that, for example, it is possible to have a new gauge 
boson in the region between 500 and 600 GeV with a coupling of 
electromagnetic strength that could be observed at Run II.
\begin{figure}  
\centerline{ \epsfysize=2.75 in \epsfbox{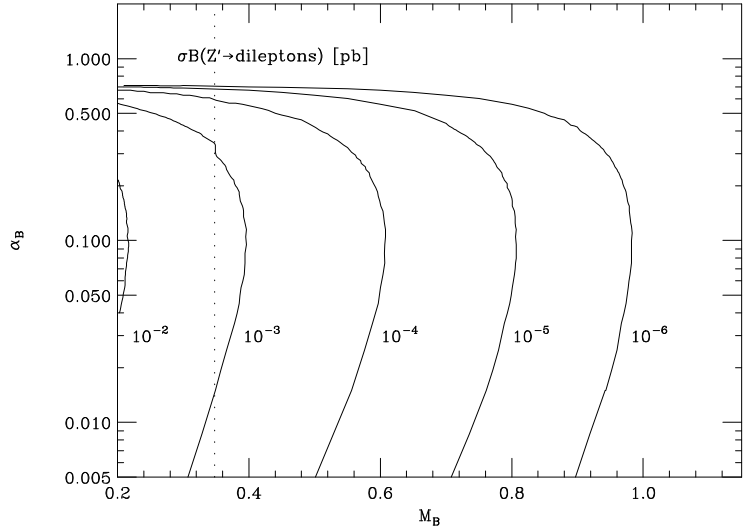}} 
\caption{Contours of constant cross section times branching fraction to
dileptons.  The dotted line shows the threshold $M_B=2 m_{top}$.}
\label{figure2}
\end{figure}

{\em Decays to Dileptons}.  Given the construction of our model,
the specification of $M_B$ and $\alpha_B$ is sufficient to determine 
the magnitude of $c_B(M_B)$, up to a small uncertainty.  For each point in 
the parameter space, $M_B/\sqrt{4\pi\alpha_B}$ is of order the scale 
of U(1)$_B$ breaking.  However, this scale also determines the masses 
of the exotic fermions, and the point at which $c_B$ begins to run.  The 
only uncertainty is in the Yukawa couplings of the exotic matter, which 
we assume is of order one (say, between $1/3$ and $3$); this only affects 
the result logarithmically.  To account for the mixing,  we use 
Eq.~(\ref{rge}) to run $c_B$ from the  U(1)$_B$ breaking scale 
$\Lambda_B=rM_B/g_B$, where $r$ is an ${\cal O}(1)$ uncertainty, down 
to $M_B$ with the condition $c_B(\Lambda_B)=0$. We show some typical 
values of $c_B$ in Table 1 for different choices of $M_B$ and $\alpha_B$.
\begin{table}[b]
\begin{center}
\begin{tabular}{ccc}
$M_B$ (TeV) & $\alpha_B(M_B)$ & $c_B(M_B)$ \\ \hline\hline
0.2 & 0.1 & 0.00688 \\
0.5 & 0.1 & 0.00694 \\
1.0 & 0.1 & 0.00699 \\
0.2 & 0.01 & 0.00469 \\
0.5 & 0.01 & 0.00471 \\
1.0 & 0.01 & 0.00473 \\ \hline
\end{tabular}
\label{table1}
\caption{Kinetic mixing for $r=3$.}
\end{center}
\end{table}
The results are uniformly small, due to two competing effects:  
if the coupling $g_B$ is reduced with $M_B$ held
fixed, then the `starting' scale $\Lambda_B$ is increased, while
the rate of running, {\em i.e.} the right-hand side of
Eq.~(\ref{rge}), is reduced.  As a consequence, the
branching fraction to leptons
\begin{equation} \label{B}
B = \frac{\frac{3}{2} c_B^2 \alpha_Y}{\frac{N_f}{9}\alpha_B +
\frac{3}{2} c_B^2 \alpha_Y} \, ,
\end{equation}
is highly suppressed throughout the parameter space in Fig.~\ref{figure1}.
Here $N_f$ is the number of quarks lighter than $M_B/2$.  In
Fig.~2 we show contours of constant $\sigma$B; note that $\sigma$B
vanishes when $\Lambda_B(\alpha_B)=M_B$.  The CDF bound
on this product in no stronger than $0.04$~pb for dilepton invariant 
masses above $\sim 400$~GeV, and is significantly weaker for smaller
masses~\cite{abe2}; as a consequence, no additional bound can be placed on 
our parameter space.  It is possible, however, that a dilepton signal
could be discerned at Run II, if the $Z'$ were already discovered 
in the dijet channel.  For example, for $M_B \approx 400$~GeV, 
where the current bound is $0.04$~pb, a simple rescaling by ${\cal L}$
suggests that the bound could become $0.0028$~pb after $20$~fb$^{-1}$
of integrated luminosity.   The results in Fig.~2 imply that
this would be sufficient to see the model's tiny dilepton signal.
\begin{figure}  
\centerline{ \epsfysize=2.75 in \epsfbox{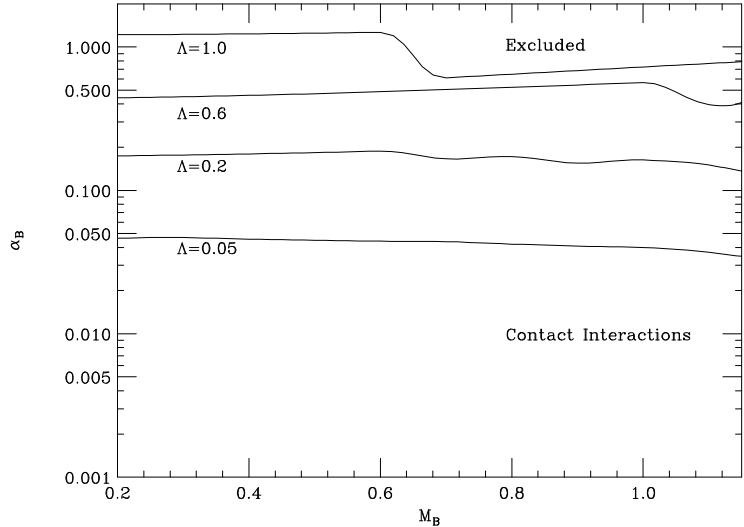}} 
\caption{Bounds obtained from the contribution of KK modes 
heavier than 1.15 TeV to
contact interactions for several values of $\Lambda$.}
\label{figure3}
\end{figure}

{\em KK-modes}. The $Z'$ phenomenology we have discussed thus far
has related to the zero-mode gauge field, and is independent of how 
the model is configured in extra dimensions.  As we mentioned earlier, 
if all the non-chiral fields propagate in the bulk, then the first 
$Z'$ KK mode is outside the reach of the Tevatron, and the zero-mode 
is of principle interest to us.  Here, we wish to consider an alternative 
possibility, that the compactification scale is low enough such that the
first few KK modes are also within the kinematic reach of the Tevatron.  
This can be the case if only U(1)$_B$ and its associated exotic Higgs 
fields live in the bulk. The usual strong bounds on $\Lambda$ are evaded
in this situation since there are {\em no} exotic Higgs fields charged
under both U(1)$_B$ and any of the standard model electroweak gauge
factors -- the vev of such a field would lead to mixing at tree-level
between the $Z$ and $Z'$ KK modes.  In order to determine the relevant
bounds, let us consider the following terms in the $Z'$ Lagrangian:
\begin{eqnarray} \label{kkcouplings}
{\cal L}_{KK} & = & -\frac{1}{4} \sum_{n=0} 
F^{(n)}_{\mu\nu} F_{(n)}^{\mu\nu}+ \frac{1}{2} \sum_{n=0} 
\left(M_B^2 + \Lambda^2 n^2 \right) Z^{\prime \mu (n)} Z_{\mu}^{\prime (n)}
\nonumber \\
&&- \frac{g_B}{3} \, \bar{q} \gamma^\mu \left( Z^{\prime (0)}_{\mu} 
+ \sqrt{2} \, \sum_{n=1} Z_{\mu}^{\prime (n)} \right) q \, .
\end{eqnarray}
Notice that the KK modes have contributions to their masses from
both the symmetry breaking and the compactifaction scale.  If
$\Lambda \ll M_B$, there is effectively a `pile-up' of states with 
masses of order $M_B$ and multiplicity $M_B/\Lambda$.  This
is one way in which low-energy bounds are enhanced.  In addition, the 
coupling of the KK modes to quarks has an extra factor of $\sqrt{2}$ 
compared to the coupling of the zero mode; this results from the field 
rescalings necessary to put the four-dimensional kinetic terms in 
canonical form, and to give the zero-mode gauge coupling its conventional 
normalization.  Hence, the appropriate dijet bound on a given KK mode may be
obtained from Fig.~\ref{figure1} by scaling down the 
exclusion limit shown by a factor of $2$.\footnote{The running
of $\alpha_B$ in the range shown in Figure~\ref{figure1} is
small, and can be neglected in this discussion.}  If $\Lambda$ is
sufficiently small, the zero mode and first few KK modes could be 
unobserved in Run~I, but discovered at Run~II.  We therefore consider whether
$\Lambda$ can be small enough for this interesting
situation to be obtained.

Aside from the KK modes that are within the reach of the Tevatron, there 
is also an infinite tower of heavier modes that are integrated out of the 
low-energy theory. Thus, the new physics manifests itself as a series
of narrow resonances, together with effective contact interactions that
lead to smoothly growing cross sections.  We may use the bounds
on four-quark contact interactions to bound the compactification scale.
If we integrate out all the modes with mass $M_B > M_{min}=1.15$~TeV (the
endpoint of the dijet invariant mass spectrum in Ref.~\cite{abe1})
we obtain operators of the form
\begin{equation} \label{4fermion}
{\cal L}_{\bar{q}q\bar{q}q} = -\sum_{n_{min}}^\infty
\frac{g_B^2}{9 M_n^2} \, \overline{q}_L \gamma_\mu q_L 
\overline{q}_L \gamma^\mu q_L + \cdots
\end{equation}
where $M_n^2=M_B^2+n^2 \Lambda^2$, and $n_{min}$ corresponds
to the first KK mode above $M_{min}$.  We show only the purely
left-handed operator, which is the one most tightly constrained
of those listed in the Review of Particle Physics~\cite{rpp00},
{\em viz.}, $\Lambda_{LL}^-(qqqq)>2.4$~TeV at 95\% C.L., with 
$\Lambda_{LL}^-(qqqq)$ defined therein.  The sum shown in
Eq.~(\ref{4fermion}) can be evaluated analytically so that the bound 
may be written as
\begin{equation}
\alpha_B < \frac{9 M_B\Lambda}{(2.4\mbox {TeV})^2} \left[
i \Psi(n_{min}-\frac{i M_B}{\Lambda})- 
i \Psi(n_{min}+\frac{i M_B}{\Lambda}) \right]^{-1}
\label{eq:contactb}
\end{equation}
where $\Psi(x)=\frac{\partial}{\partial x} [ \ln \Gamma (x) ] $ is the
digamma function.  We plot Eq.~(\ref{eq:contactb}) for several values of
$\Lambda$ in Fig.~3.  The mild steps in these contours occur each
time a KK mode becomes more massive than $M_{min}$, and is
included in the contact term.

\begin{figure}  
\centerline{ \epsfysize=2.75 in \epsfbox{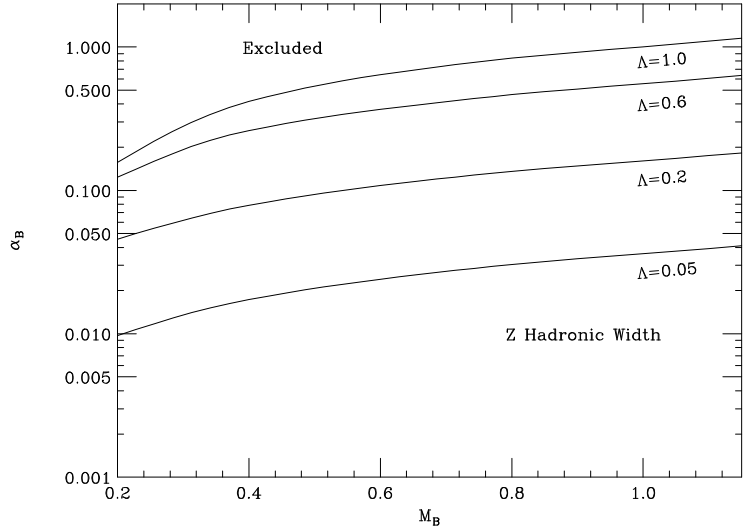}} 
\caption{Bound obtained from the contribution of the first 1000 KK 
modes to the Z hadronic width.}
\label{figure4}
\end{figure}
In the case where $\Lambda$ is small, we can also determine
whether the pile-up of states at $M_B$ is significantly bounded by 
$Z$-pole observables. The most stringent constraint for this type of 
model comes from the measurement of the $Z$ hadronic width~\cite{carmur2}, 
which is known to approximately $0.1\%$~\cite{rpp00}.  We include 
contributions from the  Z$-Z^{\prime}$ mixing~\cite{carmur2} 
and from the one-loop $q\bar{q}Z$ vertex correction~\cite{carmur1}. The 
total effect is given by
\begin{eqnarray}\label{zwidth} \nonumber
\frac{\Delta \Gamma_{had}}{\Gamma_{had}} 
  & \approx & 
-1.194 \, c_B(m_Z) \, \sqrt{\alpha_B} \, m_Z^2 \, \left(
\frac{1}{m_Z^2-M_B^2} + 2 \, \sum_{n=1}^\infty 
\frac{1}{m_Z^2-M_n^2}\right) \\
  &         & +  \frac{\alpha_B}{18 \pi} \left( F_2(M_B) + 
2 \sum_{n=1}^{\infty} F_2(M_n) \right) \, ,
\end{eqnarray}
where $c_B(m_Z)$ is found by solving Eq.~(\ref{rge}), and $F_2(M)$ is a 
loop integral factor that can be found in Ref.~\cite{carmur1}.
The sums appear linearly in Eq.~(\ref{zwidth}) since the effects
of new physics appear in an interference term at lowest order.
Figure 4 shows the 2$\sigma$ bound for several choices of $\Lambda$, 
where the sum includes the first 1000 KK modes.  Generally, the bound 
obtained from the Z hadronic width supersedes the one obtained from 
contact interactions.  Figs.~3 and 4 in conjunction with Fig.~1 show 
that the compactification scale $\Lambda$ can be made small enough
so that the $Z'$ zero mode and first few KK excitations could
be undetectable at Run I and discovered at Run II, without requiring
the coupling $\alpha_B$ to be unexplicably small.  For
example, the parameter choice $\alpha_B=0.01$, $M_B=400$~GeV,
and $\Lambda=200$~GeV is consistent with all our constraints.
    
\section{Conclusions} \label{sec:concl} \setcounter{equation}{0}

We have shown in this article that it is possible to construct
viable models with a non-anomalous U(1) symmetry that is orthogonal
to hypercharge and that preserves proton stability, a concern when the 
quantum gravity scale is low.  While exotic chiral fields are 
required to cancel anomalies, we show that these fields may
nonetheless be vector-like under the standard model subgroup, so
that constraints from the $S$ parameter are evaded, and may appear
in complete SU(5) representations, so that power-law unification
may be preserved.  The new gauge boson and its KK excitations exhibit
a high degree of leptophobia, which is only violated by kinetic
mixing with hypercharge, which is small and calculable, given
our assumed boundary conditions.  If power-law unification is sacrificed, 
then one may consider the case in which only the extra U(1)  
lives in the bulk.  In this case, the most important bounds on the 
compactification scale come from processes associated with the 
exchange of the $Z'$ and its KK excitations, and were found to be
relatively weak.  This allows the $Z'$ and its first few KK modes
to be within the kinematic reach of the Tevatron.  In both versions of
the model, we considered bounds from collider searches for new particles 
decaying to dijets and dileptons, and, in the second case, bounds on
the compactification scale from contact interactions and contributions 
to the Z hadronic width.  For gauge couplings comparable to that
of hypercharge, we showed that this scenario is allowed by current 
experiments, and that the new gauge boson, and perhaps some of 
its KK excitations could be discovered by the Tevatron at Run~II.

\begin{center}               
{\bf Acknowledgments} 
\end{center}
We thank the National Science Foundation for support under Grant No.\ 
PHY-9900657, and the Jeffress Memorial Trust for support under Grant 
No.~J-532.



\begin{thebibliography}{99} 
\frenchspacing

\bibitem{georgi}
See, for example, H. Georgi,
Ann.\ Rev.\ Nucl.\ Part.\ Sci.\  {\bf 43}, 209 (1993).

\bibitem{add}
N. Arkani-Hamed, S. Dimopoulos, G. Dvali, Phys.\ Lett.\ B429 (1998) 263;
Phys.\ Rev.\ D59 086004 (1999); I.~Antoniadis, N.~Arkani-Hamed, 
S.~Dimopoulos and G.~Dvali, Phys.\ Lett.\  {\bf B436}, 257 (1998)
[hep-ph/9804398].

\bibitem{fv}
Z.~Berezhiani and G.~Dvali,
Phys.\ Lett.\  {\bf B450}, 24 (1999)
[hep-ph/9811378].

\bibitem{kane}
F.~C.~Adams, G.~L.~Kane, M.~Mbonye and M.~J.~Perry,
hep-ph/0009154.

\bibitem{carmur2}
C.D. Carone and H. Murayama, 
Phys.\ Rev.\  {\bf D52}, 484 (1995) [hep-ph/9501220].

\bibitem{sk}
Y.~Hayato {\it et al.}  [SuperKamiokande Collaboration],
Phys.\ Rev.\ Lett.\  {\bf 83}, 1529 (1999)
[hep-ex/9904020]

\bibitem{s33}
C.~D.~Carone, L.~J.~Hall and H.~Murayama,
Phys.\ Rev.\  {\bf D53}, 6282 (1996)
[hep-ph/9512399].

\bibitem{ahs}
N.~Arkani-Hamed and M.~Schmaltz,
Phys.\ Rev.\  {\bf D61}, 033005 (2000)
[hep-ph/9903417].

\bibitem{br}
For a list of references, see
I.~Antoniadis and K.~Benakli,
Int.\ J.\ Mod.\ Phys.\  {\bf A15}, 4237 (2000)
[hep-ph/0007226].

\bibitem{carmur1}
C.D. Carone and H. Murayama, 
Phys.\ Rev.\ Lett.\  {\bf 74}, 3122 (1995)
[hep-ph/9411256].

\bibitem{dav}
D.~Bailey and S.~Davidson,
Phys.\ Lett.\  {\bf B348}, 185 (1995)
[hep-ph/9411355].

\bibitem{ac}
A.~Aranda and C.~D.~Carone,
Phys.\ Lett.\  {\bf B443}, 352 (1998)
[hep-ph/9809522].

\bibitem{aa}
I.~Antoniadis, Phys.\ Lett.\  {\bf B246}, 377 (1990);
I.~Antoniadis and K.~Benakli, Phys.\ Lett.\  {\bf B326}, 69 (1994)
[hep-th/9310151].

\bibitem{ddg}
K.~R.~Dienes, E.~Dudas and T.~Gherghetta,
Phys.\ Lett.\  {\bf B436}, 55 (1998)
[hep-ph/9803466]; Nucl.\ Phys.\  {\bf B537}, 47 (1999)
[hep-ph/9806292].

\bibitem{dkm}
K.~R.~Dienes, C.~Kolda and J.~March-Russell,
Nucl.\ Phys.\  {\bf B492}, 104 (1997)
[hep-ph/9610479].

\bibitem{erler}
J.~Erler, Nucl.\ Phys.\  {\bf B586}, 73 (2000)
[hep-ph/0006051]; H.~Cheng, B.~A.~Dobrescu and K.~T.~Matchev,
Nucl.\ Phys.\  {\bf B543}, 47 (1999)
[hep-ph/9811316]; Phys.\ Lett.\  {\bf B439}, 301 (1998)
[hep-ph/9807246].

\bibitem{gu}
C.~D.~Carone, Phys.\ Lett.\  {\bf B454}, 70 (1999)
[hep-ph/9902407]; P.~H.~Frampton and A.~Rasin, Phys.\ Lett.\  
{\bf B460}, 313 (1999) [hep-ph/9903479]; A.~Delgado and M.~Quiros,
Nucl.\ Phys.\  {\bf B559}, 235 (1999)
[hep-ph/9903400]; Z.~Kakushadze, Nucl.\ Phys.\  {\bf B548}, 205 (1999)
[hep-th/9811193]; A.~Perez-Lorenzana and R.~N.~Mohapatra, 
Nucl.\ Phys.\  {\bf B559}, 255 (1999) [hep-ph/9904504].

\bibitem{dfm}
K.~R.~Dienes, A.~E.~Faraggi and J.~March-Russell,
Nucl.\ Phys.\  {\bf B467}, 44 (1996)
[hep-th/9510223].

\bibitem{rpp00}
Review of Particle Physics, D.E. Groom {\em et al}, Eur.\ 
Phys.\ J.\ C15 (2000) 1. 

\bibitem{pem}
P.~Nath and M.~Yamaguchi, Phys.\ Rev.\  {\bf D60}, 116004 (1999)
[hep-ph/9902323]; Phys.\ Rev.\  {\bf D60}, 116006 (1999)
[hep-ph/9903298]; M.~Masip and A.~Pomarol, Phys.\ Rev.\  {\bf D60}, 
096005 (1999) [hep-ph/9902467]; A.~Strumia, Phys.\ Lett.\  {\bf B466}, 
107 (1999) [hep-ph/9906266]; R.~Casalbuoni, S.~De Curtis, D.~Dominici 
and R.~Gatto, Phys.\ Lett.\  {\bf B462}, 48 (1999)
[hep-ph/9907355]; T.~G.~Rizzo and J.~D.~Wells,
Phys.\ Rev.\  {\bf D61}, 016007 (2000) [hep-ph/9906234]; 
C.~D.~Carone, Phys.\ Rev.\  {\bf D61}, 015008 (2000)
[hep-ph/9907362].

\bibitem{bre}
M.~Bando, T.~Kugo, T.~Noguchi and K.~Yoshioka,
Phys.\ Rev.\ Lett.\ {\bf 83}, 3601 (1999)
[hep-ph/9906549].

\bibitem{abe1}
F.~Abe {\it et al.}  [CDF Collaboration],
Phys.\ Rev.\  {\bf D55}, 5263 (1997) [hep-ex/9702004].

\bibitem{pdf}
H.~Plothow-Besch, Int.\ J.\ Mod.\ Phys.\  {\bf A10}, 2901 (1995).

\bibitem{abe2}
F.~Abe {\it et al.}  [CDF Collaboration],
Phys.\ Rev.\ Lett.\  {\bf 79}, 2192 (1997).

\end{thebibliography}
\end{document}